\documentclass[10pt,aps,pre,twocolumn,final,letterpaper,superscriptaddress,longbibliography]{revtex4-2}

\usepackage[utf8]{inputenc}
\usepackage{graphicx}
\usepackage{placeins}
\usepackage{xcolor}
\providecolor{monbleu}{RGB}{76,114,176}
\usepackage{amsmath, amssymb}
\usepackage[caption=false, labelformat=empty]{subfig}
\usepackage[colorlinks, allcolors=monbleu]{hyperref}

\graphicspath{{./figures/}}

\begin{document}

\author{Leah A.~Keating}
\affiliation{Vermont Complex Systems Institute, University of Vermont, Burlington, VT 05405 USA}
\affiliation{Department of Computer Science, University of Vermont, Burlington, VT 05405 USA}

\author{Laurent H\'{e}bert-Dufresne}
\affiliation{Vermont Complex Systems Institute, University of Vermont, Burlington, VT 05405 USA}
\affiliation{Department of Computer Science, University of Vermont, Burlington, VT 05405 USA}
\affiliation{Santa Fe Institute, Santa Fe, NM 87501, USA}

\title{Loops, not groups: Long cycles are responsible for discontinuous phase transitions in higher-order network contagions}

\begin{abstract}
We study a self-consistent approach to introduce higher-order effects in a branching process model of complex contagion on clustered networks.  Our self-consistent solution allows us to distinguish multiple exposures that stem from a single transmission chain, from those occurring on cycles at the intersection of different transmission chains. We find that only the latter mechanism gives rise to a discontinuous phase transition in the size of global cascades, a defining feature of complex contagions. Group effects alone, without long cycles, produce continuous phase transitions.
\end{abstract}

\maketitle
Dynamics on higher-order networks have been widely studied recently \cite{majhi2022dynamics, bick2023higher, ferraz2024contagion, stonge2025defining}, with many papers focusing on the impact of the introduction of group dynamics into traditional models for spreading processes on networks. 
It is often claimed that the structure and effects of groups cause discontinuous phase transitions \cite{iacopini2019simplicial, battiston2025collective, battiston2021physics, stonge2022influential, ferraz2024contagion, breton2025explosive, malizia2025hyperedge,lamata2025hyperedge} but common mean-field analysis cannot actually resolve which network pathways are used by dynamical processes. In this Letter, we argue that the impact of groups on the phase transition is more nuanced. Using a branching process, we disentangle the impact of groups from the impact of loops (separate transmission chains entering the same group) and find that long cycles are essential for discontinuous phase transitions. 

Here, we deal with a class of networks, which we refer to as Newman-Miller networks \cite{newman2009random, miller2009percolation}. These networks are fully composed of independent edges, connecting pairs of nodes, and triangles (or \textit{3-cliques}), connecting groups of 3 nodes together. The networks are parametrized by a joint distribution of edge- and triangle-degree, with connections made at random. 


Newman and Miller's methods \cite{newman2009random, miller2009percolation} work well for simple contagion models which can be mapped to bond percolation such that all interactions (simple edges or edges within triangles) transmit the contagion with a \textit{constant} probability \cite{newman2002spread}. 
For complex contagions where the probability of transmission can increase for second exposures within triangles, Keating et al.~\cite{keating2022multitype, keating2023generating} used a multi-type branching process argument to accurately capture the size distribution of cascades in the subcritical regime.
To study the phase transition and the supercritical regime of complex contagion in higher-order networks, we also take a percolation approach and extend the methods of Newman and Miller. The framework could be applicable to networks with motifs beyond triangles, following the methods of Allard et al. \cite{allard2015general, allard2012exact, allard2012bond}.

Altogether, percolation has been widely used to study the critical properties of spreading models \cite{newman2002spread, kenah2007second, pastor2015epidemic, allard2023role} because it accurately approximates Susceptible-Infectious-Recovered dynamics \cite{kenah2007second} while opening a unique mathematical toolbox. Here, percolation is implemented as a breadth-first branching process where we track individual exposures to capture different higher-order effects.  


\paragraph*{\textit{Simple contagion}} In the simple contagion model, each active node has one chance to transmit to its inactive neighbours with probability $T$, before it is removed, with certainty, in the next timestep. We model this as a bond percolation process following the approach of Newman \cite{newman2009random} and Miller \cite{miller2009percolation}, each edge is considered occupied with probability $T$, in which case we assume that this edge will transmit the contagion if given the chance. The resulting cascade or contagion size is then given by the size of the connected components of the percolation network, and a supercritical cascade corresponds to a macroscopic giant component \cite{newman2002spread, kenah2007second}. 

Using Newman-Miller networks \cite{newman2009random, miller2009percolation}, which describe a network by the distribution of the number $s$ of independent edges, and the number $t$ of triangles, of a randomly selected node. The joint probability distribution of $s$ and $t$, $p_{s, t}$, can be described by the pgf
\begin{equation}
    G_0 (x, y) = \sum_{s, t=0}^{\infty}p_{s, t}x^{s}y^{t}.
\end{equation}
In contagion dynamics on a network, it is also necessary to know the distribution of the excess number of single edges and triangles if we reach a node by traversing a single edge, or by traversing an edge in a triangle. The pgf for the excess distribution of single edges and triangles by traversing an single edge is given by
\begin{align}
    G_s (x, y) &= \sum_{s, t=0}^{\infty}s p_{s, t}x^{s - 1}y^{t} =\frac{1}{\langle s \rangle}\frac{\partial}{\partial x} G_0 (x,y) ,
\end{align}
and by traversing an edge in a triangle is
\begin{align}
    G_t (x, y) &= \sum_{s, t=0}^{\infty}t p_{s, t}x^{s}y^{t-1} =\frac{1}{\langle t \rangle}\frac{\partial}{\partial y} G_0 (x,y).
\end{align}
In the examples in this letter, we consider the case where $p_{s, t}$ follows a doubly-Poisson distribution \cite{newman2009random},
\begin{equation}
    p_{s, t} = e^{-\mu}\frac{\mu^{s}}{s !}e^{-\nu}\frac{\nu^{t}}{t !},
\end{equation}
where $\mu$ is the mean number of independent edges per node, and $\nu$ is the mean number of triangles (groups) per node. For the doubly-Poisson distribution, $G_0 (x, y) = G_s (x , y) = G_t (x, y) = e^{\mu (x-1) +\nu (y-1) }$.

We aim to calculate the probability that a randomly selected node $i$ is in the giant component of the percolated network. We first calculate $u$, the probability that a randomly selected single edge of $i$ does not lead to the giant component, and the probability $v$ that a randomly selected triangle of $i$ does not lead to the giant component. There are two ways in which a single edge will not connect $i$ to the giant component (with probability $u$). (1) With probability $1-T$, transmission does not occur along the edge, or (2) transmission does occur along the edge with probability $T$, but with probability $G_s (u, v)$ the node at the other end of the edge is not in the giant component,
\begin{equation}
    u = 1-T + TG_s (u,v) .\label{eq:start_simple_size}
\end{equation}
There are four ways in which a triangle fails to connect node $i$ to the giant component (with probability $v$). 
(1) Transmission fails on both edges connecting $i$ to the other two nodes in the triangle, with probability $(1-T)^2$. 
(2) Transmission occurs on either one of two edges, the other two edges fail and the single reached neighbor is not in the giant component, with probability $2T(1 - T)^2 G_t(u, v)$. 
(3) Transmission occurs on either one of two edges, with the other failing but the third edge transmitting, and neither of the two neighbors are in the giant component, with probability $2T^2(1 - T) G_t(u, v)^2$. 
(4) Both edges from $i$ transmit and neither of the two neighbors are in the giant component, with probability $T^2 G_t(u, v)^2$. Summing across the four cases yields:

\begin{align}
    v = & (1-T)^2 +2T(1-T)^2 G_t(u,v)\nonumber\\ 
    &+ \left[ 2T^2 (1-T)+T^2 \right] G_t(u,v)^2.
\end{align}

\begin{figure}[]
    \centering
    \subfloat[]{
  \includegraphics[width=0.45\textwidth]{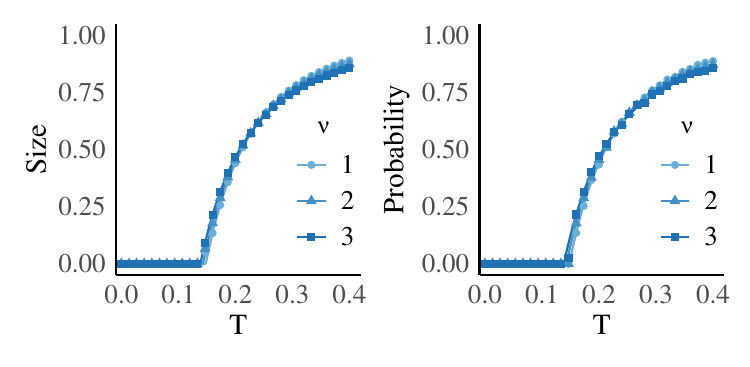}%
    }
    \caption{The average size $S$ of the largest connected component (left) and the probability $R$ ($R=S$ for simple contagion) that a node is in the largest connected component (right) for varying values of $T$, the base transmission probability. The curves are the value of $S$ from Eq.~(\ref{eq:simple_size}) for simple contagion. These results are for a doubly-Poisson distributed network with $\nu = 1$, 2, and 3 and $\mu = 6.5 - 2\nu$. The points 
    represent the median values of 10,000 simulations each on a network with at least 100,000 nodes. 
    }
    \label{fig:simple_contagion}
\end{figure}

\begin{figure}
    \centering

    \subfloat[]{
  \includegraphics[width=0.49\textwidth]{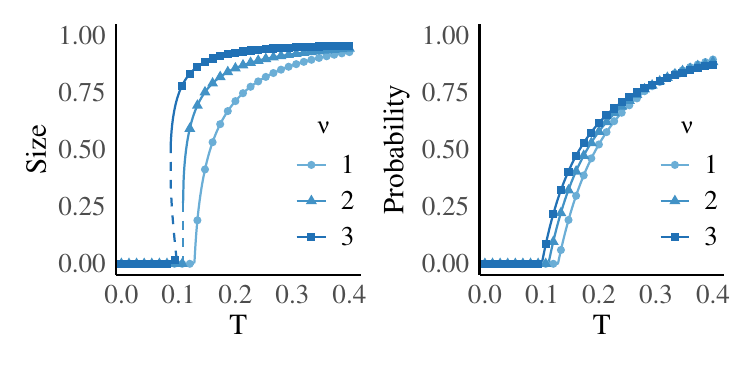}%
    }
    
    \caption{The case where $\alpha_\Delta = \alpha_\curvearrowright = 10$; i.e., groups and loops. (Left) The stable (solid lines) and unstable equilibria (dashed lines) for the largest connected component size $S$ from Eq.~(\ref{eq:loops_groups_size}) and (right) the probability $R$ of observing a supercritical cascade from Eq.~(\ref{eq:groups_loops_prob}) for a doubly-Poisson distributed network with $\nu = 1$, 2 and 3 and $\mu = 6.5 - 2\nu$ for varying values of $T$, the base transmission probability. The points
    represent the median values for 100,000 simulations each on a network with at least 100,000 nodes. The unstable equilibria were calculated as in \cite{dodds2004universal}.}
    \label{fig:loops_and_groups}
\end{figure}

\begin{figure}[t]

    \centering
 \subfloat[]{
  \includegraphics[width=0.45\textwidth]{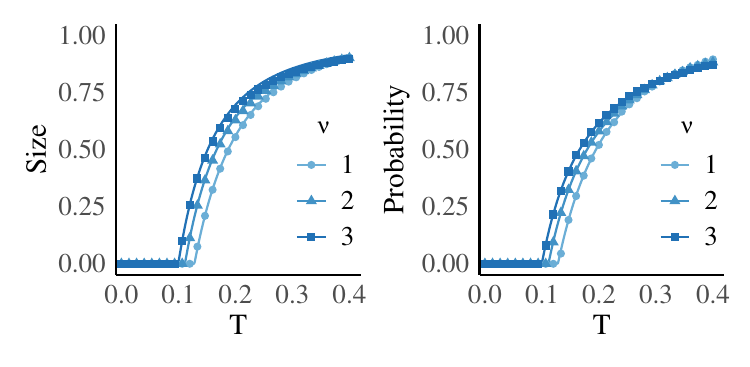}%
    }\\
    \vspace{-0.75cm}
    
    \caption{The case where $\alpha_{\Delta}=10$ and $\alpha_{\curvearrowright}=1$; i.e., groups, no loops. (Left) The largest connected component size $S$ from Eq.~(\ref{eq:loops_groups_size}) and (right) the probability $R$ of observing a supercritical cascade from Eq.~(\ref{eq:groups_loops_prob}) for a doubly-Poisson distributed network with $\nu = 1$, 2 and 3 and $\mu = 6.5 - 2\nu$ for varying values of $T$, the base transmission probability. 
    The points 
    represent the median values for 100,000 simulations each on a network with at least 100,000 nodes. 
    }
    \label{fig:groups_no_loops}
\end{figure}

\begin{figure}
    \centering
     \subfloat[]{
  \includegraphics[width=0.45\textwidth]{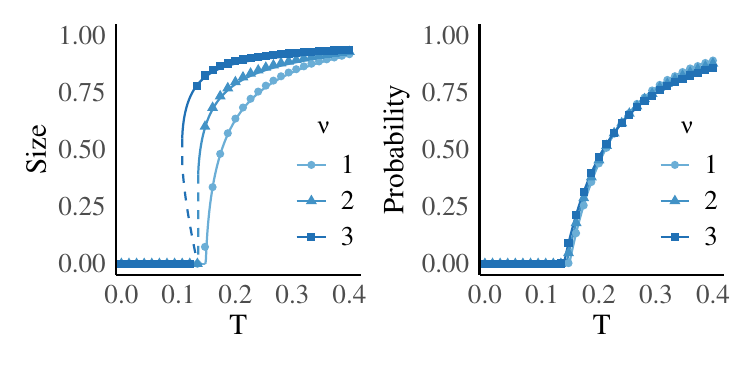}%
    }\\
    \vspace{-0.75cm}
    

    \caption{The case where $\alpha_{\Delta}=1$ and $\alpha_{\curvearrowright}=10$; i.e., loops, no groups. (Left) The stable (solid lines) and unstable equilibria (dashed lines) for the largest connected component size $S$ from Eq.~(\ref{eq:loops_groups_size}) and (right) the probability $R$ of observing a supercritical cascade from Eq.~(\ref{eq:groups_loops_prob}) for a doubly-Poisson distributed network with $\nu = 1$, 2 and 3 and $\mu = 6.5 - 2\nu$ for varying values of $T$, the base transmission probability. 
    The points 
    represent the median values for 100,000 simulations each on a network with at least 100,000 nodes. 
    }
    \label{fig:loops_no_groups}
\end{figure}

Then, the probability that a randomly selected node is \textit{not} in the giant component is $G_0 (u, v)$, and the probability $R$ that a randomly selected node is in the giant component is
\begin{equation}
    R = 1 - G_0 (u, v) .\label{eq:simple_size}
\end{equation}
For a simple contagion, Eq.~(\ref{eq:simple_size}) also gives the fractional size $S$ of the giant component \cite{newman2009random}, since the probability that an initial contagion seed is in a specific component is equal to the fractional size of the component. In Fig. \ref{fig:simple_contagion}, we show that Eq.~(\ref{eq:simple_size}) agrees well with Monte Carlo simulations (and exact in the thermodynamic limit). In the Monte Carlo simulations of Figs.~\ref{fig:simple_contagion}, \ref{fig:loops_and_groups}, \ref{fig:groups_no_loops}, and \ref{fig:loops_no_groups}, we extract the probability and size of a giant component through the fraction and median size of simulations that reach at least 1\% of nodes.

In the next section, we go beyond these simple contagion dynamics to complex contagion, where adoption can be more likely on a second exposure.

\paragraph*{\textit{Complex contagion}} We implement complex contagion as a simplicial contagion, with an increased transmission probability for a node's second exposure within a given triangle.
This increased transmission probability is
\begin{equation}
    T_\Delta = 1 - (1-T)^{\alpha_\Delta} .
\end{equation}
If $\alpha_\Delta = 1$, then $T_\Delta = T$ and we have a simple contagion within triangles, and if $\alpha_\Delta > 1$, then $T_\Delta > T$ and we have a complex contagion where second exposures within triangles have a higher transmission probability. Proceeding as in Eqs.~(\ref{eq:start_simple_size}-\ref{eq:simple_size}), and changing every second exposure to a node within a triangle to transmit with $T_\Delta$, the resulting model is 
\begin{align}
    u_{out} =& 1-T + T G_s(u_{out}, v_{out}) ,\label{eq:start_loops_groups_prob}\\
    v_{out} =& (1-T)^2 + 2T(1-T)(1-T_\Delta)G_t (u_{out}, v_{out})\nonumber\\
    &+\left[2 T T_\Delta (1-T) + T^2\right]G_t (u_{out}, v_{out})^2 ,
\end{align}
and
\begin{equation}    
    R = 1 - G_0 (u_{out}, v_{out})~,\label{eq:groups_loops_prob}
\end{equation}
where $u_{out}$ ($v_{out}$) is the probability that a node is not connected to the largest connected component through a randomly-selected edge (triangle), and $R$ is the probability that we observe a giant component. We use the subscript \textit{``out''} to refer to the direction of the percolation, where in outward percolation we consider whether a transmission link exists from a node to the giant component.

In the right panel of Fig.~\ref{fig:loops_and_groups}, the curves are the self-consistent solutions to Eqs.~(\ref{eq:start_loops_groups_prob}-\ref{eq:groups_loops_prob}), which capture the probability of a giant component, with $\nu = \{1,2,3\}$, and $\mu = 6.5-2\nu$ to fix the average degree. We compare these curves to the probability of observing a giant component in Monte Carlo simulations 
and see very good agreement. However, Eqs.~(\ref{eq:start_loops_groups_prob}-\ref{eq:groups_loops_prob}) do not capture the size of the giant component (not shown). The symmetry that exists between the size of the largest connected component and the probability of a global cascade for simple contagion is broken in complex contagion, inducing directionality.

We need to go beyond our na\"ive implementation of complex contagion to calculate the size of the giant component by considering the other direction of percolation. We define $u_{in}$ ($v_{in}$) as the probability that node $i$ selected uniformly at random is not connected to the giant component through a randomly-selected single edge (triangle), where the direction of transmission is from the other node to node $i$ (from the other two nodes to node $i$). In the \textit{inward} direction, we get that
\begin{equation}
    u_{in} = 1 -T + T G_s (u_{in}, v_{in}),\label{eq:start_loops_groups_size}
\end{equation}
which is the same as Eq.~(\ref{eq:start_loops_groups_prob}). 

\begin{figure}
    \centering
    \includegraphics[width=\linewidth]{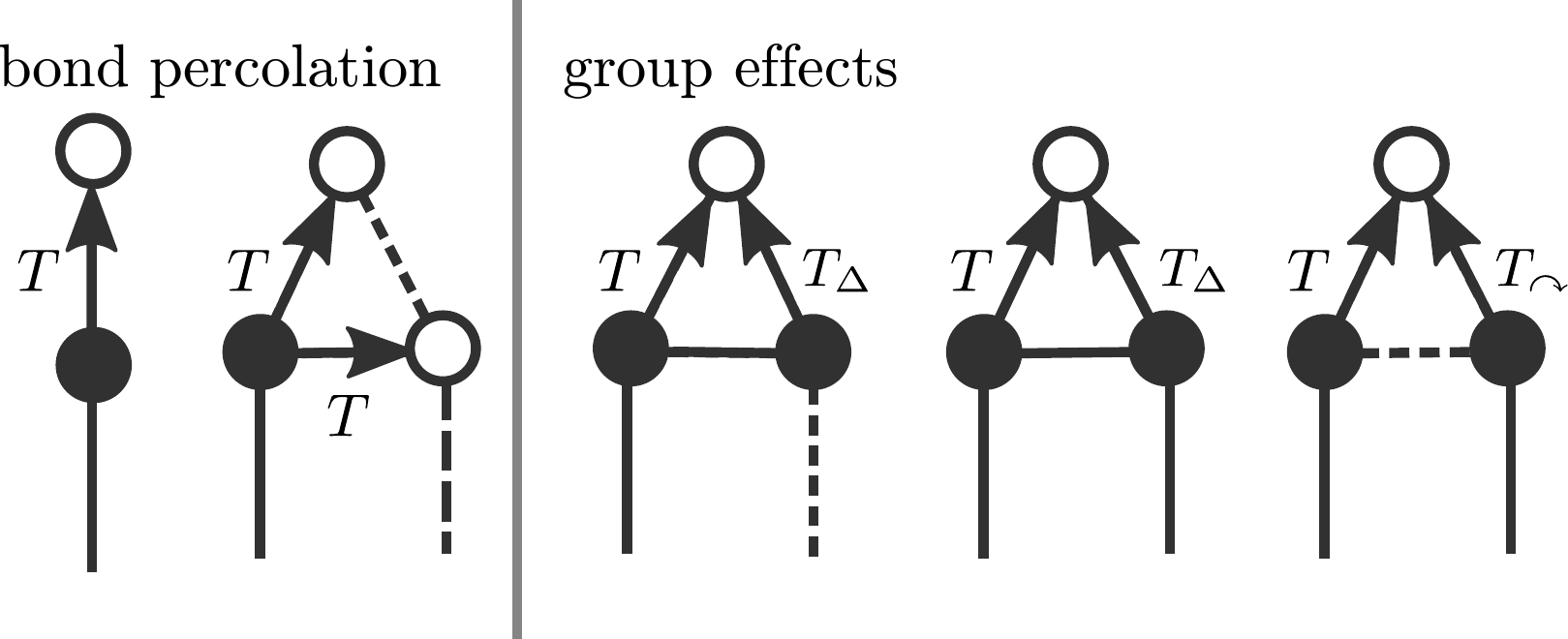}
    \caption{{\bf The mechanism of spread through a group.} (Left) The first node to be activated in a group activates its neighbours, as in classic bond percolation, with transmission probability $T$. (Right) In the situation where two nodes out of three in a triangle are active, the activation probability of the third node depends on the context of activations around the active nodes. From left to right in the group effects panel: (i) if there is an active link exists between the two active nodes in the triangle, the third node is activated by the second node with probability $T_\Delta$, (ii) the same mechanism applies with probability $T_\Delta$ even if one of the active nodes was also reached by a neighbour outside the group, and (iii) if there is no transmission link between the two active nodes in the group, implying two independent introductions in the triangle, the second exposure activates the third node with probability $T_\curvearrowright$.}
    \label{fig:network_dynamics}
\end{figure}

In a triangle, we need to count how many of the two neighbors of node $i$ are actually part of the giant component, in which case we call them activated. We find that it is necessary to consider the case where a second node can be activated in a triangle from \textit{within} the triangle, but also the case where the second node can become activated through a \textit{separate} transmission chain that also reaches the triangle (see Fig.~\ref{fig:network_dynamics}). We denote $\rho_{in}$ the probability that a randomly selected neighbor in the triangle is activated through a transmission chain external to the triangle,
\begin{equation}
    \rho_{in} = 1 - G_t (u_{in}, v_{in}).
\end{equation}

We add a further probability $T_\curvearrowright = 1 - (1-T)^{\alpha_{\curvearrowright}}$, which is the probability of transmission from a second active node in a triangle if that second node was activated through an external transmission loop. Unlike standard simplicial contagion, we now use two parameters to separate whether the second exposure is caused by a transmission within the triangle ($T_\Delta$) or from a second external introduction ($T_\curvearrowright$). Then, 
\begin{align}
    v_{in} = &  (1-T)\left[(1-T)(1-T_\curvearrowright)+T(1-T_\Delta) \right]\rho_{in}^2\nonumber\\
    & + (1-T)\left[T(1-T_\Delta) + (1-T)\right]2\rho_{in}(1-\rho_{in})\nonumber\\
    & + (1-\rho_{in})^2,
\end{align}
This equation is our key methodological contribution. The size of the largest connected component is then
\begin{equation}
        S = 1 - G_0 (u_{in}, v_{in})\label{eq:loops_groups_size} \; .
\end{equation}
The result of Eq.~(\ref{eq:loops_groups_size}) is validated in the left panel of Fig.~\ref{fig:loops_and_groups}.

\paragraph*{\textit{Distinguishing loops and groups}} We consider three cases: \textit{groups and loops} (classic complex contagion, $\alpha_\Delta = \alpha_\curvearrowright = 10$); \textit{groups, no loops} ($\alpha_\Delta = 10$ and $\alpha_\curvearrowright = 1$); and \textit{loops, no groups} ($\alpha_\Delta = 1$ and $\alpha_\curvearrowright = 10$). 

In Fig.~\ref{fig:loops_and_groups}, we show the case where $\alpha_\Delta = \alpha_\curvearrowright = 10$; i.e., groups and loops. This increased transmission probability exists regardless of whether the second active node was activated from a node inside the triangle, or through a second transmission loop. 
We observe a discontinuous phase transition in the size of the giant component, but not in the probability of existence of the giant component. Conversely, for simple contagion, in Fig.~\ref{fig:simple_contagion}, the phase transition is continuous for both the size of the giant component and the probability of existence of the giant component.

In Fig.~\ref{fig:groups_no_loops}, we show the case where $\alpha_\Delta = 10$ and $\alpha_\curvearrowright = 1$; i.e., groups, no loops. In this case, a second active node in a triangle only gets an increased probability of activating the inactive node in the triangle if it was activated itself from within the triangle ($\alpha_\Delta = 10$). 
We find phenomenology reminiscent of simple contagions with continuous phase transitions. The within-group higher-order effect $\alpha_\Delta$ increases the size and probability of existence of the giant component, but does not alone change the nature of the phase transition.

In Fig.~\ref{fig:loops_no_groups}, we show the case where $\alpha_\Delta = 1$ and $\alpha_\curvearrowright = 10$; i.e., loops, no groups. In this case, if the second active node in a triangle is activated from outside of the triangle, then it is given a higher probability of transmitting to the inactive node in the triangle ($\alpha_\curvearrowright = 10)$; but not if the second active node is activated from within the triangle ($\alpha_\Delta = 1$). We observe a discontinuous phase transition in the size of the giant component, matching the traditional complex contagion case. The higher-order effect along cycles, $\alpha_\curvearrowright$, not only increases the size and probability of existence of the giant component, but can alone change the nature of the phase transition.

\begin{figure}
    \centering
    \includegraphics[width=\linewidth]{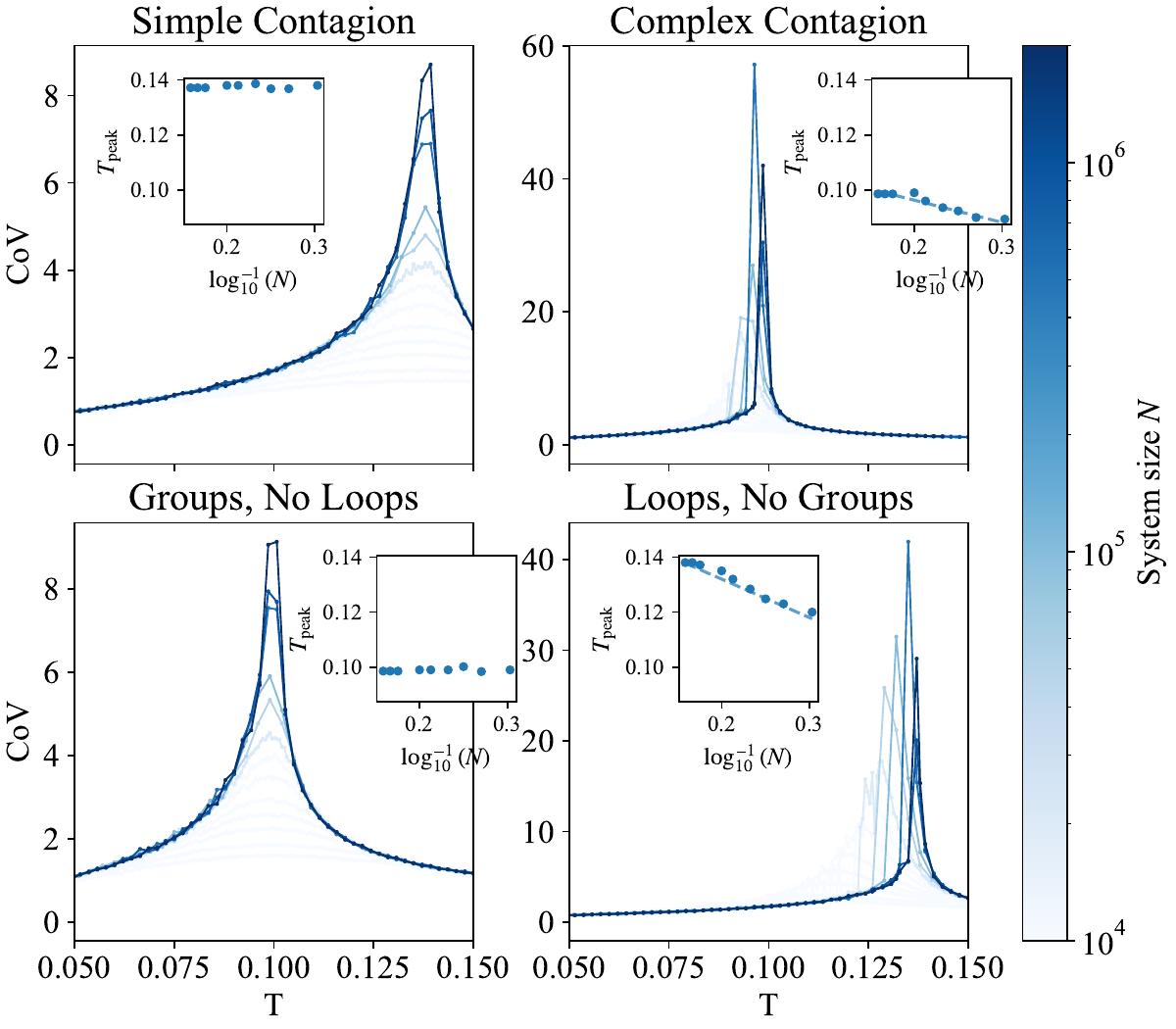}
    \caption{The coefficient of variation (CoV), which is the standard deviation divided by the mean, against the transmission parameter $T$, for all four combinations of $\alpha_\Delta, \alpha_\curvearrowright \in \{1,10\}$. Both panels on the left side have $\alpha_\curvearrowright = 1$ and have similar CoV curves always increasing with $N$ -- a sign of a continuous phase transition. Both panels on the right side have $\alpha_\curvearrowright = 10$ and have similar CoV curves with peak position shifting with $N$ but saturating in height -- a sign of a discontinuous phase transition. All points are the result of at least 15,000 simulations on networks of sizes; 2,000,000, 1,000,000, 500,000, 100,000, 50,000, 20,000, 10,000, 5,000, and 2,000.}
    \label{fig:cov_T}
\end{figure}

\paragraph*{\textit{Conclusions}} We have shown that for models of complex contagion on networks with group structure, the existence of a discontinuous phase transition in the size of the giant component depends on the role of transmission cycles (or long loops) in the network. In our model, we only observe discontinuous phase transitions in cases where cycles increase transmission probability, $T_\curvearrowright > T$ ($\alpha_\curvearrowright > 1$). We show evidence of these discontinuous phase transitions for $\alpha_\curvearrowright = 10$ in Fig.~\ref{fig:cov_T}, where we plot the Coefficient of Variation (CoV) against $T$ as a proxy for susceptibility. 
The case with $\alpha_\curvearrowright=1$ is found to mirror simple contagion where susceptibility diverges with system size, whereas $\alpha_\curvearrowright=10$ mirrors complex contagion where susceptibility saturates.
The insets in Fig.~\ref{fig:cov_T} track the critical point $T_{peak}$ as we vary system size. With $\alpha_\curvearrowright = 10$, $T_{peak}$ has a finite size correction that scales with the inverse of the expected length of the shortest cycle \cite{bollobas2011random}, i.e. $1/\log(N)$; supporting the critical role of cycles.
This result implies that features of hypergraph that hinder (promote) cycles \cite{malizia2025hyperedge}, such as overlap between hyperedges, are also likely to hinder (promote) discontinuous transitions.
Understanding the interplay of structure and dynamics is important in the growing area of higher-order networks, where it is often claimed that the mere existence of groups can cause discontinuous phase transitions.

\paragraph*{\textit{Acknowledgments}}
The authors acknowledge David JP O'Sullivan for help conceptualizing the research. The authors also acknowledge financial support from the University of Vermont's Office of the Vice President for Research and from The National Science Foundation award \#2419733. This collaboration was supported by International Partnerships and Programs at the University of Vermont as a Global Catalyst Research Partnership Award.

\bibliography{refs.bib}

\end{document}